\documentclass{aa}  

\usepackage{graphicx}
\usepackage{booktabs}
\usepackage{txfonts}
\begin{document}

   \title{Efficiency of particle acceleration at interplanetary shocks: Statistical study of STEREO observations}


   \author{N. Dresing
          \inst{1}
          \and
          S. Theesen\inst{1}
          \and
          A. Klassen\inst{1}
          \and
          B. Heber\inst{1}
          }

   \institute{Institut f\"ur Experimentelle und Angewandte Physik, University of Kiel, Germany\\
   \email{dresing@physik.uni-kiel.de} 
             }

   \date{}

 
  \abstract
   {Among others, shocks are known to be accelerators of energetic charged particles. However, many questions regarding the acceleration efficiency and the required conditions are not fully understood. In particular, the acceleration of electrons by shocks is often questioned.}
   {In this study we determine the efficiency of interplanetary shocks for $<$100\,keV electrons, and for ions at $\sim$0.1 and $\sim$2\,MeV energies, as measured by the Solar Electron and Proton Telescope (SEPT) instruments aboard the twin Solar
Terrestrial Relations Observatory (STEREO) spacecraft.}
   {We employ an online STEREO in situ shock catalog that lists all shocks observed between 2007 and mid 2014 (observed by STEREO~A) and until end of 2013 (observed by STEREO~B). In total 475 shocks are listed. To determine the particle acceleration efficiency of these shocks, we analyze the associated intensity increases (shock spikes) during the shock crossings.
   For the near-relativistic electrons, we take into account the issue of possible ion contamination in the SEPT instrument.}
   {The highest acceleration efficiency is found for low energy ions (0.1\,MeV), which show a shock-associated increase at 27\% of all shocks. The 2\,MeV ions show an associated increase only during 5\% of the shock crossings. In the case of the  electrons, the shocks are nearly ineffective. Only five shock-associated electron increases were found, which correspond to only 1\% of all shock crossings.}
   {}

   \keywords{Interplanetary shocks, shock acceleration, electron acceleration}

   \maketitle
%

\section{Introduction}
Collisionless shocks are known to be able to accelerate energetic charged particles inside the heliosphere \citep[e.g.,][]{Tsurutani1985}, within the solar corona \citep[e.g.,][]{Mann2005}, at the Earth's bow shock \citep[e.g.,][]{Burgess2007}, while, in other astrophysical systems \citep[e.g.][]{Blandford1987, Amato2006}, shocks are powerful accelerators for producing  Galactic cosmic rays (GCRs) (see \cite{Takamoto2015} and references therein). 
The heliosphere contains various shocks, such as planetary bow shocks as well as propagating shock fronts like those that arise in stream interaction regions (SIRs) or are driven by coronal mass ejections (CMEs). 
Large gradual solar energetic particle (SEP) events \citep{Reames1999}, which show long lasting proton enhancements, are associated with CME-driven shocks.
This source mechanism is usually not associated with electron-rich events and the role of coronal and interplanetary (IP) shocks for electron acceleration is poorly understood, compared to that of ions (see \cite{Pomoell2015} and references therein).
While modeling results show that highly relativistic electrons, whose gyroradii are close to the gyroradii of ions, can be accelerated to high energies by dif\/fusive shock acceleration, the acceleration of low-rigidity electrons is dif\/ficult to compute via this mechanism.
The gyro radii of these electrons are too small to resonantly interact with magnetic fluctuations in the shock region \citep[][and references therein]{Guo2010} which is referred to as the injection problem.
Nevertheless, in situ observations in the heliosphere show several examples of electron acceleration at shocks \citep{Gosling1989, Decker2005, Simnett2005}.
In these studies the electrons were found to be accelerated at quasi-perpendicular shocks,  where the shock drift acceleration is very efficient for electron acceleration up to supra-thermal energies \citep{Holman1983, Krauss-Verban1989}.
However, recent modeling approaches incorporating strong large-scale magnetic fluctuations suggest that electrons can be ef\/ficiently accelerated, regardless of the shock orientation \citep{Guo2015}.\\
While particles accelerated at a remote shock may propagate to a magnetically connected observer, usually a peak in the ion intensities is observed when a shock passes a spacecraft in the heliosphere.
This peak is called the shock spike and is believed to represent the locally accelerated particles.
In situ observations at spacecraft have, however, shown that not every interplanetary shock accelerates particles. 
Many studies have focused on the acceleration efficiency of protons at in situ shocks.
\cite{Kallenrode1996} found shock-associated increases for 5\,MeV protons in 53\% of the analyzed shocks using Helios data, \cite{Huttunen2005} have determined an efficiency of 48\% for $>$1.5\,MeV protons that were detected during seven years of SOlar and Heliospheric Observatory (SOHO) observations using the Energetic and Relativistic Nuclei and Electron (ERNE) instrument.
Using measurements from the International Sun Earth Explorer-3 (ISEE 3) of 37 shocks, \cite{Tsurutani1985} found a lower efficiency of 27\% for 1.5\,MeV protons. 
They also point out that the acceleration efficiency is higher for lower energies which is confirmed by other authors \citep[e.g.,][]{Lario2003, Ho2008}.
This is understood to be an effect of the shock-drift acceleration, which is less efficient for particles that can quickly escape from the shock owing to their higher energy.
Using ACE/EPAM data \cite{Lario2003} determine efficiencies of 61\% and 33\% for 57\,keV and 3\,MeV ions, respectively.\\
\cite{Tsurutani1985} and \cite{Lario2003} also analyzed the response of energetic electrons at shock crossings.
For $>$2\,keV electrons \cite{Tsurutani1985} found an efficiency of 68\%.
However, at higher energies, they only find 8\% ($>$20\,keV) and 0\% (60\,keV electrons), respectively.
\cite{Lario2003} find an efficiency of 17\% for 45\,keV electrons.
We note that the results of the two studies are not in agreement with each other. 
The number of electron events observed by \cite{Lario2003} have at least twice the number found by \cite{Tsurutani1985}.
To shed light on this discrepancy we used the large data set of the two Solar Terrestrial Relations Observatory (STEREO) spacecraft which were launched in late 2006 and since then perform Earth-like orbits around the Sun.
They have built an observational platform in the ecliptic at $\sim$1\,AU, taking measurements that cover the rising and maximum phase of solar cycle 24 since 2007.
To determine the acceleration efficiency of shocks for electrons and ions, we investigate the STEREO/SEPT data (see Section \ref{sec:instr}) of $<$100\,keV electrons and 0.1 and 1\,MeV ions, respectively, during 475 shock crossings.
Since our main focus is on the acceleration efficiency for electrons, we have analyzed the electron data critically with respect to a potential proton contamination.
This effect is described in detail in the paper by \cite{Muller-Mellin2007} and is shown in Figure 15 of that paper using GEANT 4 simulations.
See the next section for details.
In Section \ref{data_selection} we describe the data and our selection criteria.
The results are summarized in Section \ref{results}.
Finally we discuss the results in Section \ref{discussion} and compare them to a similar study by \cite{Lario2003} using ACE/EPAM data.
%
%
%
\section{Instrumentation}\label{sec:instr}
\begin{figure}
\centering
\includegraphics[width=0.5\textwidth, clip=true, trim = 0mm 0mm 0mm 0mm]{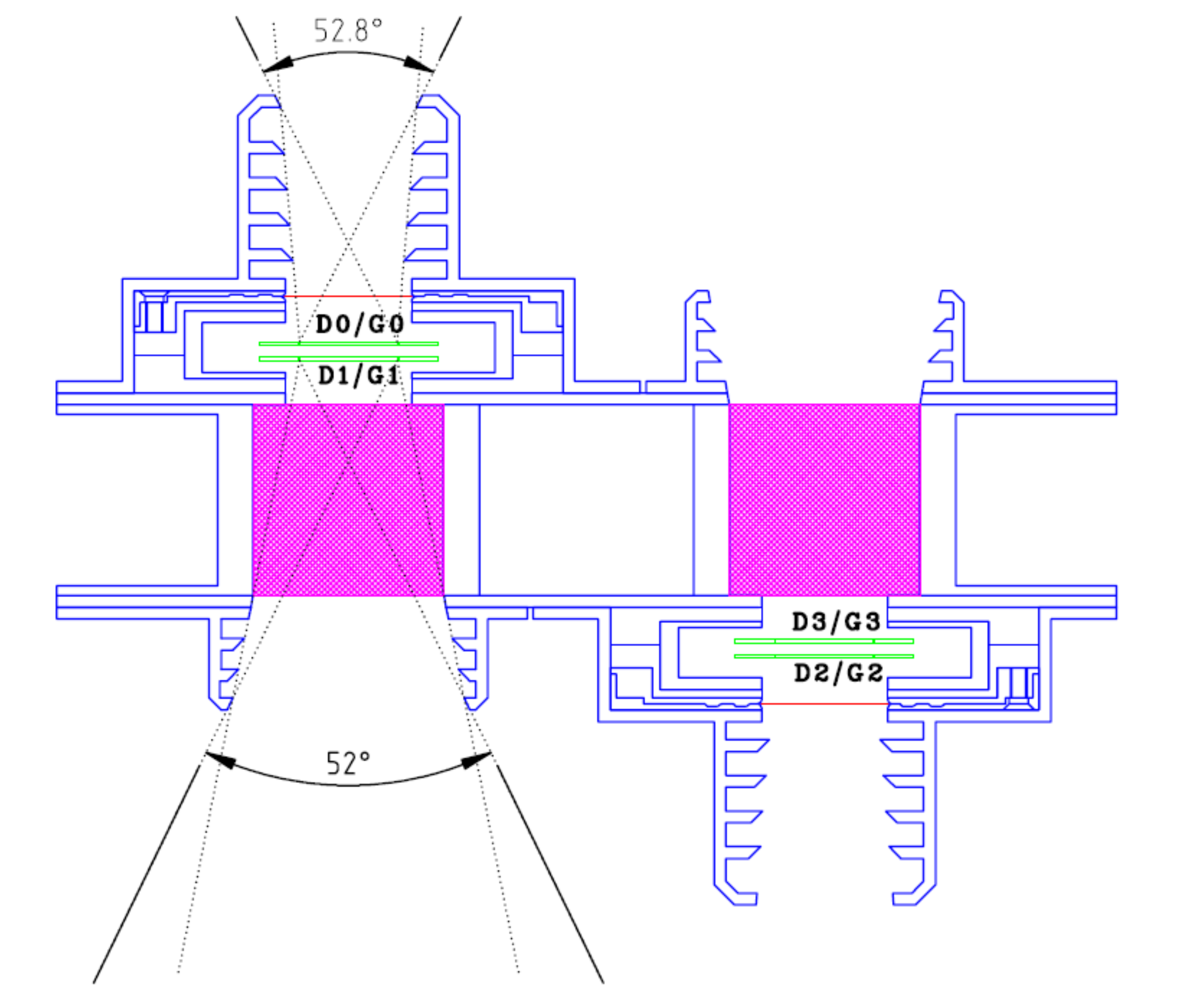}
  \caption{Schematic of the SEPT instrument. Two equal double-ended telescope units are mounted oppositely to measure electrons and ions from both directions. The side covered by the Parylene foil (thin red line) builds the electron telescope while the side with the magnet (magenta block) is dedicated to ion measurements. The two solid state detectors in the center of the units (green lines) are operated in anti-coincidence.} \label{fig:sept}
\end{figure}

The Solar Electron and Proton Telescope (SEPT, \cite{Muller-Mellin2007}) aboard the STEREO spacecraft measures electrons and ions using the magnet-foil technique.
Figure \ref{fig:sept} shows a schematic of the instrument.
It consists of two equal double-ended units that are mounted oppositely.
The two silicon detectors are indicated by D0/G0 and D1/G1, respectively.
Both detectors consist of an inner part (D0 and D1) and an outer guard ring (G0 and G1) and are operated in anti-coincidence i.e., only a signal in the inner segment of one of the two detectors leads to a valid coincidence. 
One side of the double-ended telescope is equipped with a foil (thin red line) which stops protons with energies $<$400\,keV and the other side holds a magnet (magenta block) which sweeps away $<$400\,keV electrons before they can be detected by the silicon detector.
Consequently one side of the unit builds the electron telescope providing an energy range from $\sim$50\,keV to 400\,keV and the other side builds the ion telescope with an energy range of 50\,keV to 7\,MeV.
By combining both units, electrons and protons can be detected from both directions.
Each STEREO spacecraft carries two of these SEPT telescopes so that four dif\/ferent viewing directions are covered.
One unit looks along the nominal angle of the Parker spiral at 45$^{\circ}$ towards and away from the Sun. 
The second unit is mounted in the North/South direction. \\
%
%
The work by \cite{Muller-Mellin2007} clearly shows that protons with energies between 450 and 650\,keV contaminate the SEPT electron channels.
The same simulations show that, above 400\,keV, electrons may also contribute to the ion channels. 
Therefore the intensity time profiles of electrons, as well as ions, need to be reviewed critically. 
A first order preliminary correction of the electrons has been developed utilizing two ion channels $>$390\,keV, with one bin representing proton contamination (see Fig. 15 of \cite{Muller-Mellin2007}) and the other bin representing helium contamination at energies of about four times that of the proton bin.
The method developed so far does not take into account the spectral shape of the proton spectrum as well as the abundance of helium, which is measured by the ion channel around 1.6\,MeV.
Therefore, we compare the uncorrected and corrected rates to decide if the increase is due to electrons or due to ions that are counted in the telescope. 
In Section \ref{results} we show that the number of events, where ions contaminate the electron channels, is not negligible.
%
In the upper panel of Figure \ref{real_event:49.1}, we see the ten minute averaged intensity time profiles of 102-110\,keV,  389-409\,keV, and 1.98-2.22\,MeV ions and, in the lower panel, the corresponding intensity time profiles of 65-75\,keV electrons. 
The red curve in the lower panel gives the corrected profiles. 
In this specific event, the electron correction does not significantly alter the intensity time profiles.  
\\
We note that in Section \ref{conclusion} of this study we compare the electron measurement by the SEPT instrument with a similar analysis performed with the Electron Proton Alpha Monitor (EPAM, \cite{Gold1998}) aboard the ACE spacecraft.
The EPAM instrument also uses the magnet/foil detection technique.
However, EPAM's Low Energy Magnetic Spectrometer (LEMS) telescopes provides the so-called deflected electron (DE) channels, which are believed to be free of ion contamination.


\section{Data selection and analysis}\label{data_selection}
\begin{figure}
\centering
\includegraphics[width=0.5\textwidth, clip=true, trim = 8mm 0mm 8mm 0mm]{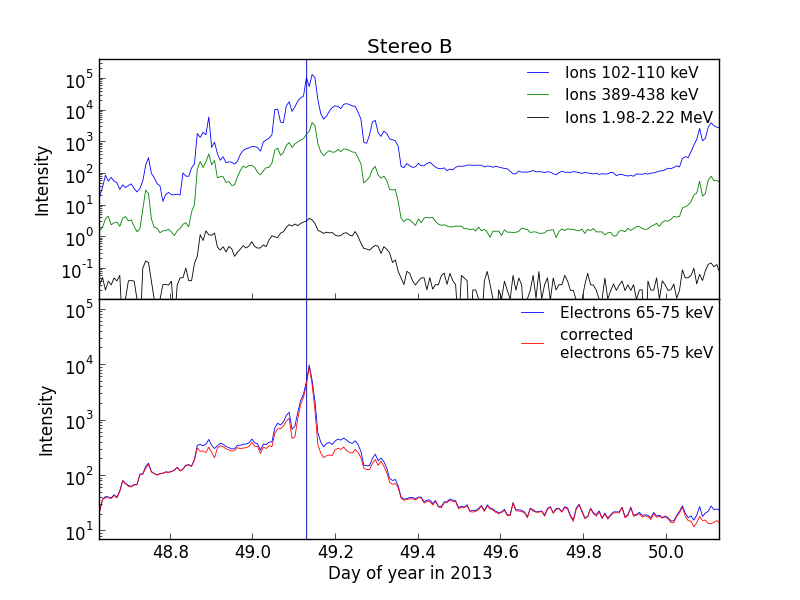}
  \caption{Ion and electron intensities around the shock crossing on 18 Feb 2013, observed by STEREO~B. The upper panel shows ions in the energy ranges of 102-110\,keV (blue), 389-438\,keV (green), and 1.98-2.22\,MeV (black), the lower panel shows electron intensities in the range of 65-75\,keV (blue), as well as the corresponding corrected electrons (red). The vertical line indicates the time of the shock passage over the spacecraft. The electrons show a clear spike-like increase at the shock.} 
  \label{real_event:49.1}
\end{figure}

\begin{figure}
\centering
\includegraphics[width=0.5\textwidth, clip=true, trim = 8mm 0mm 8mm 0mm]{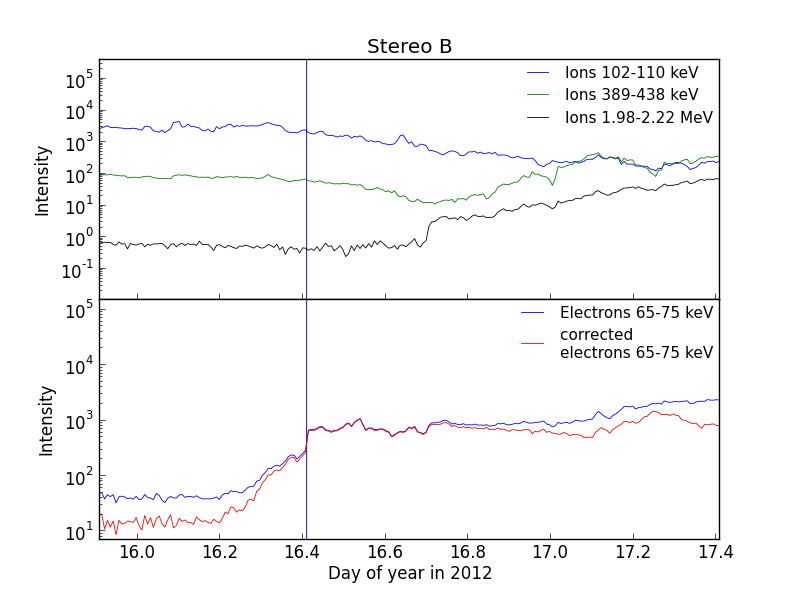}
  \caption{Ion and electron intensities, as in Fig. \ref{real_event:49.1}, for the shock crossing on 16 Jan 2012. While no shock-associated ion increase is observed, the electrons show a step-like increase which occurs on top of a rising flank.} 
  \label{real_event:16.4}
\end{figure}

%
%
We used the STEREO in situ shock lists \citep{Jian2013} provided at \url{http://www-ssc.igpp.ucla.edu/forms/stereo/stereo_level_3.html}.
These lists include the time of each shock passage at the spacecraft, as well as further characteristics of the shock such as its type (forward or reverse shock), the magnetosonic Mach number, and of the shock-normal angle.
Furthermore, the lists provide the source of the shock, such as a stream interaction region (SIR) or an interplanetary coronal mass injection (ICME).
From the comments on the source given in the lists, we identified the following four source types: SIR, ICME, SIR/ICME (both phrases "SIR" and "ICME" appeared in the comment), and other (none of the phrases "SIR" or "ICME" appeared).
At the time of our analysis the shock list for STEREO~A (STA) covered the period from 2007 until mid of 2014, and the list for STEREO~B (STB) from 2007 until end of 2013, respectively.
For each of the  475 shocks (269 at STA and 206 at STB), we analyzed electron and ion intensities that were measured by the SEPT instruments during the time of the shock passage to determine if an associated shock spike, i.e., locally accelerated particles, was observed.
We focus on near-relativistic electrons in the energy range of 65-75\,keV, and ions in the range of 102-110\,keV, and 1.98-2.22\,MeV.
Regarding the $\sim$100\,keV ions, we cross-checked the selected events in terms of a coincidence with upstream and magnetospheric bursts of the Earth's magnetosphere, which were observed during 2007 when the STEREO spacecraft were still close to the Earth \citep{Klassen2008, Muller-Mellin2008}.
Figures \ref{real_event:49.1} and~\ref{real_event:16.4} show two example events with ion intensities in the top panel and electron intensities in the lower panel. 
The ion energy channel at 389-438\,keV (green line) is the energy bin most relevant to ion contamination in the electron channels.
The lower panel shows uncorrected (blue) and corrected electron (red) time profiles, as described in Section \ref{sec:instr}.
The vertical line in Figures \ref{real_event:49.1} and~\ref{real_event:16.4} marks the time of the shock crossing.
To identify a shock-associated ion increase, the intensity around the shock passage time was determined as a 20-minute average and compared to a quiet-time background value that was determined in early 2007 when the SIR and solar activity were very low.
If the intensity value around the shock was higher than that value, the 4h pre-event background was determined and again compared to the intensity at the shock passage.
If the intensity was higher than this pre-event background by a factor of two the event was listed as a shock-associated increase.
This has been done for two different ion energy channels at $\sim$100\,keV and at $\sim$2\,MeV.
In the case of electrons, the issue of ion contamination  had to be taken into account to identify a real shock-associated electron increase.
Therefore, after identifying an increase above quiet-time and above pre-event background, the increase had to be inspected in terms of ion contamination.
For this purpose, we additionally require that the contamination is less than 60\%, i.e., the ratio of the corrected-to-measured electrons is larger than 0.4.
Finally, the evaluated candidates for shock-accelerated electron events were cross-checked in terms of a coincidence with a solar energetic particle (SEP) event:
If, by chance, SEPs arrive at the spacecraft at the same time as a shock passes the spacecraft they would be wrongly counted as shock-accelerated particles.
Therefore, radio- and solar-image observations have been analyzed to identify possible solar activity which causes the observed  electron increase.
\\
%
%
%
%
\section{Results}\label{results}
\begin{figure}
\centering
\includegraphics[width=0.5\textwidth, clip=true, trim = 0mm 0mm 0mm 0mm]{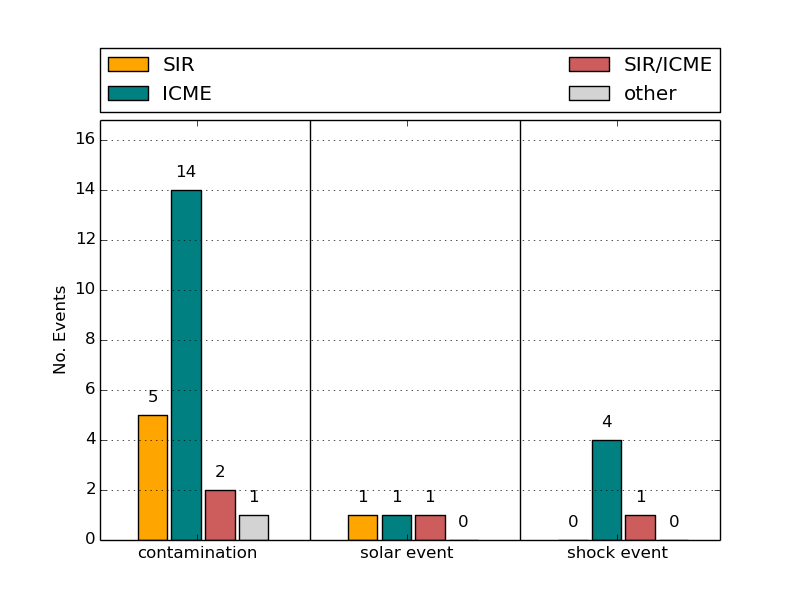}
  \caption{Histogram of all 65-75\,keV electron increases during in situ shock crossings divided into cases of strong ion contamination (left), increases due to a solar event (middle), and real shock-accelerated increases (right). The chart furthermore divides the numbers as a function of the different shock sources: SIR (yellow), ICME (green), SIR/ICME mixture (red), and "other", not clearly defined sources (gray).} 
  \label{bar_chart:electron}
\end{figure}

\begin{figure}
\centering
\includegraphics[width=0.5\textwidth, clip=true, trim = 0mm 0mm 0mm 0mm]{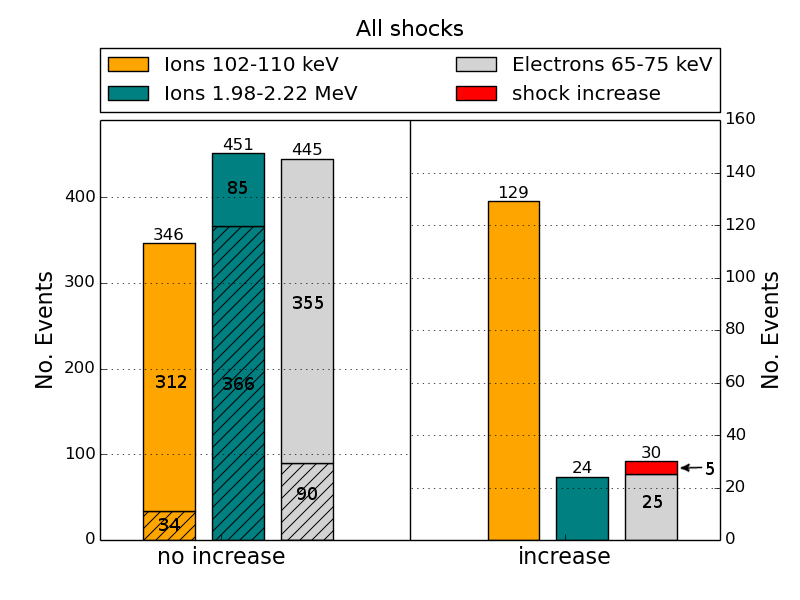}
  \caption{Histogram of all shock crossings. The yellow, green, and gray bars represent 102-110\,keV ions, 1.98-2.22\,MeV ions, and  65-75\,keV electrons, respectively. The left panel shows the number of shock crossings without a shock-associated increase. The lower parts of the bars (hatched) denote the events where only quiet-time background was measured, the upper parts of the bars represent periods of enhanced intensities which were, however, not due to the shock. The right panel shows the number of shock-associated increases. The bar representing electron increases is separated into a lower gray part, which shows the number of increases which were contaminated by ions or as a result of a solar event, and the upper red part, which represents the real shock-associated electron events.}
  \label{bar_chart:all}
\end{figure}

Table \ref{table:summarize} summarizes the analysis of near-relativistic electron increases and lists the number of shock crossings in the different categories according to our selection criteria. 
From top to bottom we list the total number of shocks, and events where the intensity stays at quiet-time background level, no increase is observed above pre-event background, the increase is not reliable owing to ion contamination, the increase occurs by accident owing to a solar event, and finally the number of real shock-associated electron increases (shock event). 
As the STEREO in situ shock lists provide the source of the shocks we separated our results accordingly (see Table \ref{table:summarize}).
In total 475 shocks crossings were analyzed. 
The results for near-relativistic electron acceleration are as follows:
\begin{itemize}
\item During 90 shock crossings (19\%) the electron intensity stayed at quiet-time background, during a further 355 crossings (75\%) no electron increase above an enhanced pre-event background was observed.
\item 30 (6\%) shock crossings where accompanied by an increase but 22 of these were caused by ion contamination, so that 73\% of the detected electron increases are contaminated and likely not real.
\item Only eight real electron increases (2\%) could be identified, of them three were related to a solar event. Therefore, only five shock-associated electron events (1\%) were observed in our sample.
\end{itemize}
\begin{table}[h!]
  \centering
    \caption{Number of shock crossings on both STEREO spacecraft divided according to the different shock sources (from left to right) and associated 65-75\,keV electron increases. From top to bottom, the numbers represent all shock crossings, no increase above quiet-time background, no increase above a 4h pre-event background, an increase that is due to ion contamination, an increase due to a solar event, and a real shock-associated increase.}
  \begin{tabular}{llllll}
  \toprule
    & ICME & SIR & SIR/ & Other & All \\
    &  &  & ICME &  &  \\
    \midrule
    Total number  & 209 & 193  & 23 & 50 & 475 (100\%)\\
    Background    & 16 & 61 & 4 & 9 & 90 (19\%)\\
    No increase   & 174 & 126 & 15 & 40 & 355 (75\%)\\
    Contamination & 14 & 5 & 2 & 1 & 22 (5\%)\\
    Solar event   & 1 & 1 & 1 & 0 & 3 (1\%)\\
    Shock event    & 4 & 0 & 1 & 0 & 5 (1\%)\\
    \bottomrule
  \end{tabular}
  \label{table:summarize}
\end{table}
Figure \ref{bar_chart:electron} shows a histogram of all the observed electron increases that were observed during the shock crossings, in order of shock source.
The left, middle, and right panels summarize the increases caused by ion contamination, events which were identified as solar events, and the remaining real shock-associated electron increases, respectively.
The majority of increases are associated with ICME-driven shocks (green bars).
The dates and shock parameters of the five real shock-accelerated electron events are listed in Table \ref{table:five_events}.
\begin{table}[b!]
  \centering
    \caption{Parameters of the five shocks, which caused a 65-75\,keV electron increases: Date, observing spacecraft (s/c), type, Mach number (M), and shock-normal angle $\theta_{Bn}$. The last column shows if the peak of the shock spike was observed upstream (up) or downstream (down) of the shock}
  \begin{tabular}{llllll}
  \toprule
    Date & S/c & type & M & $\theta_{Bn}$ & Up/down\\
    \midrule
    21 Mar 2011 & STB & ICME     & 1.25 & 80.4$^{\circ}$ & Down \\    
    16 Jan 2012 & STB & ICME     & 1.81 & 31.0$^{\circ}$ & Down \\
     7 May 2012 & STB & SIR/ICME & 1.41 & 85.6$^{\circ}$ & Unclear \\
    28 May 2012 & STA & ICME     & 2.85 & 74.8$^{\circ}$ & Up \\    
        18 Feb 2013 & STB & ICME     & 2.39 & 84.6$^{\circ}$ & Down \\
        \bottomrule
  \end{tabular}
  \label{table:five_events}
\end{table}
Four of them were observed at an ICME-driven shock and one at a SIR/ICME-mixed source shock.
The Mach numbers vary between 1.25 and 2.85.
The shock-normal angles show that four of the events were associated with a quasi-perpendicular shock and only one event with a quasi-parallel shock.
The last column in Table \ref{table:five_events} shows if the peak of the associated electron spike was observed upstream (up) or downstream (down) of the shock.
Figures \ref{real_event:49.1} and~\ref{real_event:16.4} show the two different shapes we observed during the five real shock-associated events.
Four of the cases show a peaked increase (shock spike) similar to the event observed on 18 Feb 2013 (Fig. \ref{real_event:49.1}).
Also the ions show a small peak-like increase, however the increase in the electrons is much higher so that ion contamination is not an issue here.
Fig. \ref{real_event:16.4} is the only example for a step-like increase in our sample.
The increase is detected on top of a rising flank that is associated with a solar event. 
In this case the ions do not show a shock-associated increase. \\
Figure \ref{bar_chart:all} shows a histogram where the results for electrons (gray bars) and the two selected ion channels at $\sim$ 100\,keV (yellow bars) and $\sim$ 2\,MeV (green bars) are summarized.
In this figure all shock crossings, regardless of its source, were combined.
The left panel presents the number of shock crossings without a shock-associated increase and the right panel shows the numbers of shock-associated increases.
The bars in the left panel are separated into two parts where the lower hatched part represents shock crossings accompanied by quiet-time background (i.e., instrumental background caused by cosmic rays) and the upper part shows the number of shock crossings during periods of an enhanced background level.
This enhanced background is caused by previous solar energetic particle events, which fill the inner heliosphere with low energy particles. 
As the solar activity increases the number of shock crossings during quiet-time background periods decreases (see Table \ref{table:all}).
%
%
Note, that a very small shock-associated increase might not be detected during periods of enhanced pre-event background while the detection during quiet-time background is more certain.
The electron bar in the right panel is divided into real shock-associated electron increases (red) and increases either due to solar events or increases caused by ion contamination (gray).
We note that the bars on the right-hand side are not divided into events during quiet-time or during increased background.
However, all five electron events (red bar) occurred during periods of enhanced pre-event background.
Obviously, the acceleration efficiency is larger for the ions in the keV range compared to the MeV range. 
The results for the acceleration efficiency of protons are as follows:
\begin{itemize}
\item At 102-110\,keV, we observed 129 shock-associated increases (27\%).
\item The ions in the 1.98-2.22\,MeV range only show an associated increase during 24 shock crossings (5\%).
\end{itemize}
With five real shock-associated electron increases (1\%) the electrons show the least acceleration efficiency.\\
Figures \ref{bar_chart:ICME} to \ref{bar_chart:other} show the same numbers as Fig. \ref{bar_chart:all}, but separated by the different sources of the shock: Fig. \ref{bar_chart:ICME} presents the numbers for ICME-driven shocks, Fig. \ref{bar_chart:SIR} represents SIR-associated shocks, Fig. \ref{bar_chart:SIR_ICME} shows shocks of mixed SIR-ICME events, and Fig. \ref{bar_chart:other} presents all events where the source was not clearly identified in the STEREO in situ shock lists.
Tables \ref{table:all} to \ref{table:other} show the counting numbers in detail for each year and for STA and STB separately, as well as for all shocks together and for the different source types separately.
The bottom part of each table shows the mean Mach numbers and shock-normal angles for each category.
When combining all shocks (Table \ref{table:all}) a slight dependence on Mach number can be identified:
The highest mean Mach numbers are observed for the cases that show shock spikes.
Also the shock-normal angels are largest for the cases with shock-associated increases, which suggests that these events may be associated with the most quasi-perpendicular shocks of the sample.

%
%
\section{Discussion}\label{discussion}
The results of our statistical analysis of in situ shock crossings and associated energetic ion and electron increases showed that the IP shocks are rather ineffective in $>$65\,keV electron acceleration. 
In only 1\% of the shock crossings was an associated intensity increase (shock spike)  observed in the 65-75\,keV electron energy bin.
Ions at $\sim$2\,MeV also showed a very moderate response at the shock crossings.
In only 5\% of all shocks an increase above the background was  observed.
Although the ions in the $\sim$100\,keV range were most efficiently accelerated by in situ shocks, the response of 27\% is still relatively small.\\
In earlier studies, \cite{Tsurutani1985} and \cite{Lario2003} investigated energetic ion and electron increases at 37 and 168 ICME-driven shocks that were observed by ISEE 3 and ACE, respectively.
\cite{Lario2003} separated these events according to the shape of the time profiles into the following classes: nothing (no increase), energetic storm particle (ESP) event, spike, ESP+spike, step-like, and irregular. 
To make a quantitative comparison between our analysis and the study by \cite{Lario2003}, we combine all of their classes that show shock-associated increases regardless of the time-profile shapes and only consider increases in our sample which are associated with ICME-driven shocks.
We further exclude the  "irregular" class of the \cite{Lario2003} study because our selection criteria would not detect those events.
Table \ref{table:lario_vergleich} shows a comparison of the numbers found in this study and the ones by \cite{Lario2003} and by \cite{Tsurutani1985}.
We note that the selected STEREO/SEPT energy channels are comparable to the ACE/EPAM  and ISEE 3 channels.
\begin{table}[h!]
  \centering
    \caption{Comparison of shock-associated energetic ion and electron increases found in this study (only for ICME-associated shocks) and in the studies by \protect{\cite{Lario2003}} (irregular cases excluded) and by \protect{\cite{Tsurutani1985}}. The third column shows the number of shock-associated electron increases after the events caused by ion contamination have been excluded.}
  \begin{tabular}{llll}
  \toprule
    & No  &   & No ion  \\
    & increase &  Increase &  contam. \\
        \midrule
    {\bf Ions keV} &  &  & \\
    \quad This work 102-110\,keV       & 128 & 81 (39\%) & \\
    \quad Lario et al. 47-68\,keV      & 65  & 65 (50\%) & \\    
    \quad Tsurutani \& Lin $>$47\,keV  & 10  & 27 (73\%)  &  \\
\\
    {\bf Ions MeV} &  &  & \\
    \quad This work 1.98-2.22\,MeV    & 189 & 20 (10\%) & \\
    \quad Lario et al. 1.9-4.8\,MeV   & 113 & 38 (25\%) & \\    
    \quad Tsurutani \& Lin 1.5\,MeV   & 27  & 10 (27\%) & \\
    \\
    {\bf Electrons} & &  & \\
    \quad This work 65-75\,keV & 190 & 18 (9\%) & 4 (2\%)\\
    \quad Lario et al. 38-53\,keV & 140 & 24 (15\%) & \\
    \quad Tsurutani \& Lin $>$20\,keV& 34 & 3 (8\%) & \\
    \quad Tsurutani \& Lin $>$60\,keV & 37 & 0 (0\%) & \\
    \bottomrule
  \end{tabular}
  \label{table:lario_vergleich}
\end{table}

In common to all three studies is the fact that the acceleration efficiency is energy dependent, decreasing from lower to higher energies.
While the efficiency for protons at about 1.5\,MeV found by \cite{Tsurutani1985} is in good agreement with the study by \cite{Lario2003}, our study indicates a significantly lower acceleration efficiency.
This may be due to the different analysis periods. 
Our study only considers the period during solar cycle 24, which is associated with weaker solar activity, and lower interplanetary magnetic field and density \citep{Wang2009}.
 The results are vice versa for electrons.
The acceleration efficiency of our study agrees very well with the one for $>$60\,keV electrons found by \cite{Tsurutani1985}.
The value of 15\% found by \cite{Lario2003} is neither in agreement with the one found by \cite{Tsurutani1985} at even lower energies nor with the value found in this study.
However, the value is more comparable to our number of electron increases (9\%) before excluding those events that were caused by ion contamination, which suggests that this effect may also be an issue for the ACE/EPAM LEMS telescope.\\
Although our study yields only a very small acceleration efficiency for $>$65\,keV electrons, the five found increases show that electron acceleration by shocks in the solar wind is not impossible.
The presence of these rare events raises the question of which special conditions or (shock-) characteristics must be present for an efficient electron acceleration.
To determine if a very strong shock or a certain shock type such as a quasi-perpendicular shock are the main ingredients, Table \ref{table:five_events} summarizes those values.
The mean magnetosonic Mach number of all 475 analyzed shocks is 1.6, showing that no clear association with very strong shocks can be found in our five events that have varying Mach numbers from 1.25 to 2.85.
Furthermore, only four of our five events are associated with a quasi-perpendicular shock, while one event occurs at a quasi-parallel shock.
Nevertheless, in agreement with the study by \cite{Tsurutani1985}, all of the spike-like electron increases (four out of the five events) are associated with quasi-perpendicular shocks.
We note that the majority of electron increases analyzed by \cite{Tsurutani1985} were, however, step-like post-shock increases, which did not show any dependence on the shock-normal angle.
Table \ref{table:five_events} also shows if the maximum of the electron increase was observed upstream or downstream of the shock which, however, does not provide a clear trend as well.
In contrast to that, all five events occurred during an increased intensity background, which suggests that an enhanced seed population might be important for electron acceleration.
%

%
\section{Summary and Conclusions}\label{conclusion}
We analyzed 475 interplanetary shock crossings that were observed by the two STEREO spacecraft in terms of locally shock-accelerated electrons and ions (shock spikes) and detected with the SEPT instruments.
We took into account all the shocks listed in the in situ shock catalogs provided at \url{http://www-ssc.igpp.ucla.edu/forms/stereo/stereo_level_3.html} regardless of their source such as SIRs or ICMEs.
We find that only 27\% (129 of 475 events) of the shocks are accompanied by an ion shock spike in the energy range of 102-110\,keV.
The acceleration efficiency of 5\% is even less for 1.98-2.22\,MeV ions.
The near-relativistic electrons at 65-75\,keV only show a shock-associated increase in 1\% of  cases (five events).
However, we find 25 further cases when the electron intensities increase owing to contamination from $\geq$400\,keV protons, which corresponds to 5\% of all analyzed shocks (see Fig. \ref{bar_chart:electron}). 
Thus the number of shock-associated electron increases may be larger than five.
Out of the five shock-associated electron increases, four were observed during an ICME-driven shock and one was observed at a SIR forward shock with an ICME embedded in the SIR.
Four of these events were associated with quasi perpendicular shocks and one event to a quasi parallel shock (shock-normal angle of 31$^\circ$).
Even though  five events cannot  yield statistically significant conclusions on the required parameters needed for electron acceleration, the shock parameters of the five events such as the magnetososnic Mach number or the shock normal angle did not show a clear trend (see Table \ref{table:five_events}).
The only condition that all five events had in common was the presence of an enhanced pre-event background, which could be a hint of the importance of a seed population.
%

\begin{acknowledgements}
We acknowledge the STEREO PLASTIC, IMPACT, and SECCHI teams for providing the data used in this paper. 
The STEREO/SEPT Chandra/EPHIN and SOHO/EPHIN project is supported under grant 50OC1302 by the Federal Ministry of Economics and Technology on the basis of a decision by the German Bundestag.
\end{acknowledgements}

\bibliographystyle{aa.bst}
\bibliography{references}


%
\begin{appendix}
\section{Histrograms for each of the different shock sources}

\begin{figure}
\centering
\includegraphics[width=0.5\textwidth, clip=true, trim = 0mm 0mm 0mm 0mm]{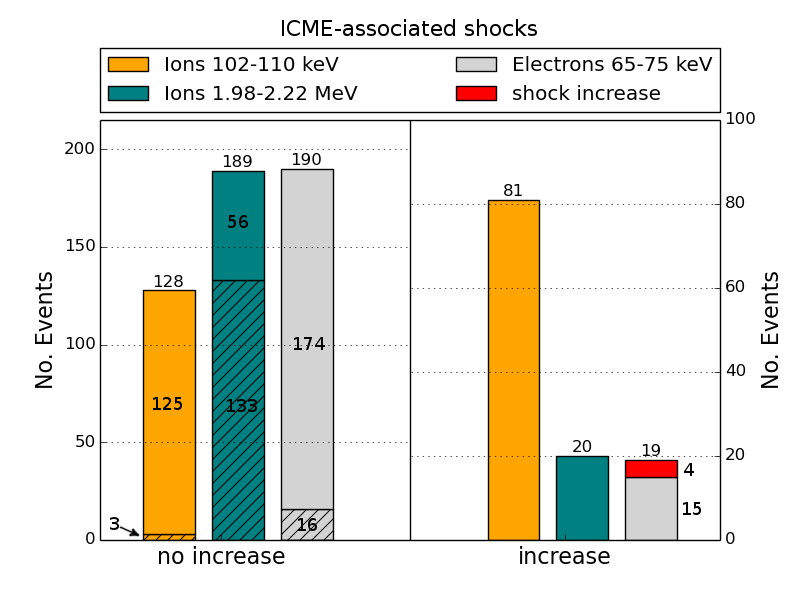}
  \caption{Histogram as in Fig. \ref{bar_chart:all} but only for ICME-associated shocks.}
  \label{bar_chart:ICME}
\end{figure}%
\begin{figure}
\centering
\includegraphics[width=0.5\textwidth, clip=true, trim = 0mm 0mm 0mm 0mm]{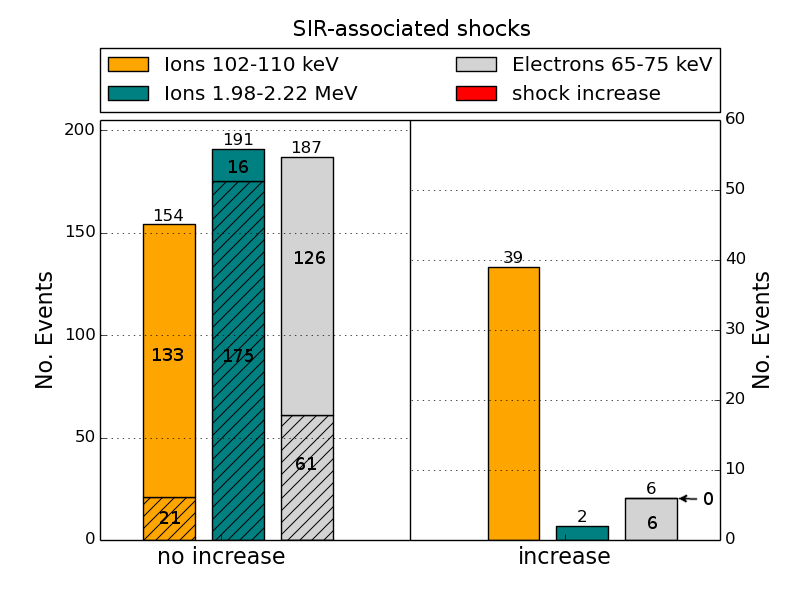}
  \caption{Histogram as in Fig. \ref{bar_chart:all} but only for SIR-associated shocks.} 
  \label{bar_chart:SIR}
\end{figure}%
\begin{figure}
\centering
\includegraphics[width=0.5\textwidth, clip=true, trim = 0mm 0mm 0mm 0mm]{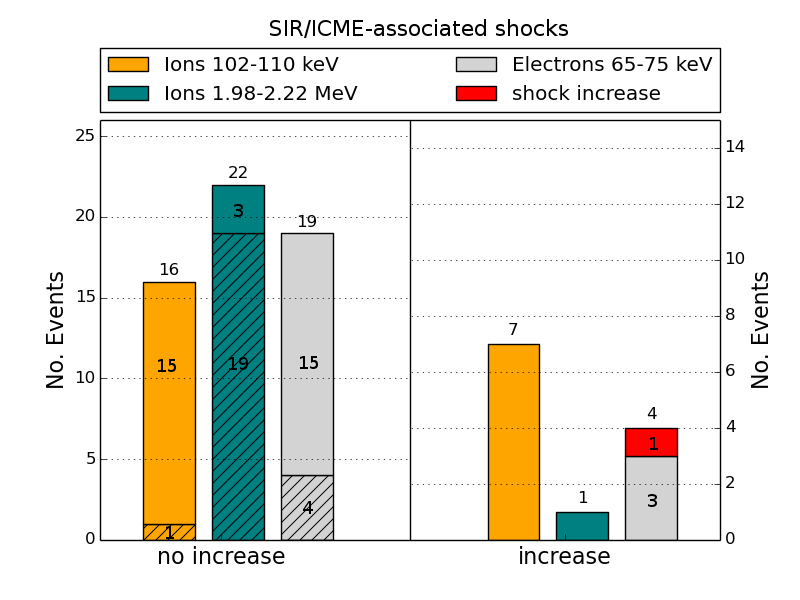}
  \caption{Histogram as in Fig. \ref{bar_chart:all} but only for shocks where SIR and ICME was mentioned in the source comments of the shock lists.} 
  \label{bar_chart:SIR_ICME}
\end{figure}%
\begin{figure}[h!]
\centering
\includegraphics[width=0.5\textwidth, clip=true, trim = 0mm 0mm 0mm 0mm]{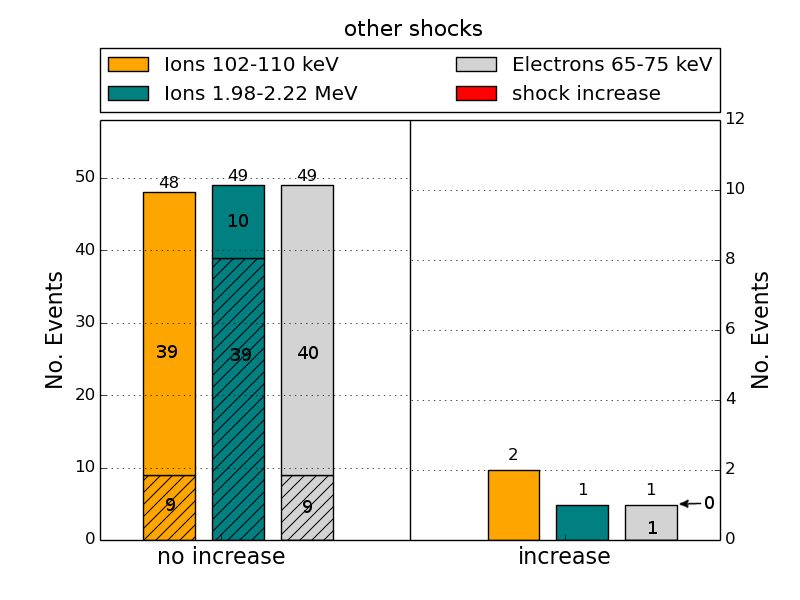}
  \caption{Histogram as in Fig. \ref{bar_chart:all} for all events of the shock lists where no clear source was specified.} 
  \label{bar_chart:other}
\end{figure}%

%
%
\section{Tables of the detailed counting numbers}
\begin{table*}[ht!]
\centering
\caption{Numbers of all shock crossings and corresponding ion and electron measurements at STEREO~A (STA) and STEREO~B (STB) for the years 2007 to mid 2014 (STA) or end of 2013 (STB). For each group, the columns show how many shock crossings were accompanied by quiet-time background, an enhanced pre-event background (but no shock-associated increase), and a shock-associated increase. For electrons it also lists how many shock crossings were accompanied by ion contamination, which makes the shock-spike identification unreliable. The bottom part summarizes the numbers for the two STEREO spacecraft together and presents the averaged Mach number M and the average shock-normal angle $\theta_{Bn}$ of all associated shocks.}
\label{table:all}
\small
\begin{tabular}{@{}lllllllllll@{}}
\textbf{All shocks}  \\
\toprule
      & \textbf{Ions keV}      &            &            & \textbf{Ions MeV}   &            &            & \multicolumn{2}{l}{\textbf{Electrons keV} } &              &            \\
       \cmidrule(l){2-4}  \cmidrule(l){5-7}   \cmidrule(l){8-11}  
           & Back-   & No shock-  & Shock-     & Back-    & No shock-  & Shock-     & Back-    & No shock-  & Contami- & Shock-     \\
                &  ground & associated & associated &  ground  & associated & associated &  ground  & associated & nation   & associated \\
                &               & increase   & increase   &            & increase   & increase   &               & increase   &          & increase   \\
  \midrule
\textbf{STA}    & & & & & & & & & & \\
2007     & 2    & 13   & 2    & 17   & 0    & 0    & 11   & 6    & 0    & 0    \\
2008     & 9    & 9    & 3    & 21   & 0    & 0    & 7    & 13   & 1    & 0    \\
2009     & 1    & 14   & 1    & 16   & 0    & 0    & 12   & 3    & 1    & 0    \\
2010     & 1    & 10   & 8    & 18   & 1    & 0    & 6    & 12   & 1    & 0    \\
2011     & 0    & 32   & 16   & 32   & 11   & 5    & 1    & 44   & 2    & 1    \\
2012     & 2    & 46   & 12   & 36   & 19   & 5    & 1    & 56   & 2    & 1    \\
2013     & 1    & 38   & 18   & 44   & 10   & 3    & 1    & 51   & 4    & 1    \\
2014     & 0    & 23   & 8    & 20   & 10   & 1    & 0    & 31   & 0    & 0    \\
sum      & 16   & 185  & 68   & 204  & 51   & 14   & 39   & 216  & 11   & 3    \\
         &      &      &      &      &      &      &      &      &      &      \\
\textbf{STB}    &      &      &      &      &      &      &      &      &      &      \\
2007     & 2    & 13   & 5    & 19   & 0    & 1    & 14   & 5    & 1    & 0    \\
2008     & 2    & 6    & 7    & 14   & 1    & 0    & 6    & 9    & 0    & 0    \\
2009     & 5    & 11   & 4    & 20   & 0    & 0    & 19   & 1    & 0    & 0    \\
2010     & 4    & 16   & 5    & 22   & 2    & 1    & 6    & 17   & 2    & 0    \\
2011     & 2    & 25   & 9    & 25   & 9    & 2    & 2    & 30   & 3    & 1    \\
2012     & 2    & 25   & 18   & 29   & 11   & 5    & 0    & 38   & 4    & 3    \\
2013     & 1    & 31   & 13   & 33   & 11   & 1    & 4    & 39   & 1    & 1    \\
sum      & 18   & 127  & 61   & 162  & 34   & 10   & 51   & 139  & 11   & 5    \\
         &      &      &      &      &      &      &      &      &      &      \\
sum A\&B & 34   & 312  & 129  & 366  & 85   & 24   & 90   & 355  & 22   & 8    \\
av. M    & 1.39 & 1.52 & 1.80 & 1.54 & 1.63 & 2.23 & 1.48 & 1.58 & 1.96 & 1.98 \\
av. $\theta_{Bn}$ & 55.1 & 58.2 & 65.5 & 60.7 & 54.7 & 67.9 & 59.4 & 59.7 & 73.6 & 67.7 \\
\bottomrule
\normalsize
\end{tabular}
\end{table*}

\begin{table*}[]
\centering
\caption{Same as in Table \ref{table:ICME}, but only for ICME-associated shocks.}
\label{table:ICME}
\small
\begin{tabular}{@{}lllllllllll@{}}
\textbf{ICME shocks}  \\
\toprule
      & \textbf{Ions keV}      &            &            & \textbf{Ions MeV}   &            &            & \multicolumn{2}{l}{\textbf{Electrons keV} } &              &            \\
       \cmidrule(l){2-4}  \cmidrule(l){5-7}   \cmidrule(l){8-11}  
           & Back-   & No shock-  & Shock-     & Back-    & No shock-  & Shock-     & Back-    & No shock-  & Contami- & Shock-     \\
                &  ground & associated & associated &  ground  & associated & associated &  ground  & associated & nation   & associated \\
                &               & increase   & increase   &            & increase   & increase   &               & increase   &          & increase   \\
  \midrule
\textbf{STA}    & & & & & & & & & & \\
2007            & 0             & 1          & 0          & 1          & 0          & 0          & 0             & 1          & 0        & 0          \\
2008            & 1             & 1          & 0          & 2          & 0          & 0          & 1             & 1          & 0        & 0          \\
2009            & 0             & 3          & 1          & 4          & 0          & 0          & 3             & 1          & 0        & 0          \\
2010            & 0             & 5          & 4          & 8          & 1          & 0          & 2             & 7          & 0        & 0          \\
2011            & 0             & 13         & 10         & 14         & 6          & 3          & 1             & 21         & 1        & 0          \\
2012            & 0             & 24         & 10         & 17         & 13         & 4          & 0             & 31         & 2        & 1          \\
2013            & 0             & 13         & 15         & 19         & 6          & 3          & 0             & 25         & 3        & 0          \\
2014            & 0             & 13         & 7          & 14         & 5          & 1          & 0             & 20         & 0        & 0          \\
sum             & 1             & 73         & 47         & 79         & 31         & 11         & 7             & 107        & 6        & 1          \\
                &               &            &            &            &            &            &               &            &          &            \\
\textbf{STB}    &               &            &            &            &            &            &               &            &          &            \\
2007            & 1             & 1          & 0          & 2          & 0          & 0          & 1             & 1          & 0        & 0          \\
2008            & 0             & 1          & 1          & 2          & 0          & 0          & 0             & 2          & 0        & 0          \\
2009            & 1             & 3          & 3          & 7          & 0          & 0          & 7             & 0          & 0        & 0          \\
2010            & 0             & 5          & 1          & 3          & 2          & 1          & 1             & 5          & 0        & 0          \\
2011            & 0             & 14         & 6          & 13         & 5          & 2          & 0             & 16         & 3        & 1          \\
2012            & 0             & 14         & 12         & 13         & 8          & 5          & 0             & 20         & 4        & 2          \\
2013            & 0             & 14         & 11         & 14         & 10         & 1          & 0             & 23         & 1        & 1          \\
sum             & 2             & 52         & 34         & 54         & 25         & 9          & 9             & 67         & 8        & 4          \\
                &               &            &            &            &            &            &               &            &          &            \\
sum A\&B        & 3             & 125        & 81         & 133        & 56         & 20         & 16            & 174        & 14       & 5          \\
av. M & 1.44          & 1.62       & 1.87       & 1.65       & 1.69       & 2.28       & 1.51          & 1.71       & 1.90     & 1.90       \\
av. $\theta_{Bn}$  & 51.8          & 57.2       & 67.5       & 61.9       & 56.1       & 70.1       & 61.78         & 59.6       & 75.7     & 70.4  \\
\bottomrule  
\normalsize
\end{tabular}
\end{table*}

\begin{table*}[]
\centering
\caption{Same as in Table \ref{table:ICME}, but only for SIR-associated shocks.}
\label{table:SIR}
\small
\begin{tabular}{@{}lllllllllll@{}}
\textbf{SIR shocks}  \\
\toprule
      & \textbf{Ions keV}      &            &            & \textbf{Ions MeV}   &            &            & \multicolumn{2}{l}{\textbf{Electrons keV} } &              &            \\
       \cmidrule(l){2-4}  \cmidrule(l){5-7}   \cmidrule(l){8-11}  
           & Back-   & No shock-  & Shock-     &Back-    & No shock-  & Shock-     & Back-    & No shock-  & Contami- & Shock-     \\
                &  ground & associated & associated &  ground  & associated & associated &  ground  & associated & nation   & associated \\
                &               & increase   & increase   &            & increase   & increase   &               & increase   &          & increase   \\
  \midrule
\textbf{STA}    & & & & & & & & & & \\
2007         & 2    & 11   & 2    & 15   & 0    & 0    & 10   & 5    & 0    & 0    \\
2008         & 6    & 8    & 3    & 17   & 0    & 0    & 5    & 11   & 1    & 0    \\
2009         & 0    & 9    & 0    & 9    & 0    & 0    & 7    & 2    & 0    & 0    \\
2010         & 1    & 4    & 3    & 8    & 0    & 0    & 4    & 4    & 0    & 0    \\
2011         & 0    & 16   & 5    & 16   & 4    & 1    & 0    & 19   & 1    & 1    \\
2012         & 0    & 10   & 1    & 8    & 3    & 0    & 0    & 11   & 0    & 0    \\
2013         & 1    & 15   & 3    & 17   & 2    & 0    & 1    & 17   & 1    & 0    \\
2014         & 0    & 3    & 0    & 2    & 1    & 0    & 0    & 3    & 0    & 0    \\
sum          & 10   & 76   & 17   & 92   & 10   & 1    & 27   & 72   & 3    & 1    \\
         &      &      &      &      &      &      &      &      &      &      \\
\textbf{STB} &      &      &      &      &      &      &      &      &      &      \\
2007         & 1    & 10   & 5    & 15   & 0    & 1    & 11   & 4    & 1    & 0    \\
2008         & 2    & 4    & 6    & 11   & 1    & 0    & 6    & 6    & 0    & 0    \\
2009         & 2    & 7    & 1    & 10   & 0    & 0    & 9    & 1    & 0    & 0    \\
2010         & 2    & 10   & 3    & 15   & 0    & 0    & 4    & 10   & 1    & 0    \\
2011         & 1    & 6    & 2    & 7    & 2    & 0    & 1    & 8    & 0    & 0    \\
2012         & 2    & 8    & 3    & 10   & 3    & 0    & 0    & 13   & 0    & 0    \\
2013         & 1    & 12   & 2    & 15   & 0    & 0    & 3    & 12   & 0    & 0    \\
sum          & 11   & 57   & 22   & 83   & 6    & 1    & 34   & 54   & 2    & 0    \\
             &      &      &      &      &      &      &      &      &      &      \\
sum A\&B     & 21   & 133  & 39   & 175  & 16   & 2    & 61   & 126  & 5    & 1    \\
av. M        & 1.37 & 1.49 & 1.61 & 1.49 & 1.57 & 2.67 & 1.47 & 1.50 & 1.69 & 3.5  \\
av. $\theta_{Bn}$   & 56.9 & 59.0 & 61.8 & 59.7 & 58.5 & 41.7 & 60.6 & 58.5 & 74.9 & 20.7 \\
\bottomrule
\normalsize
\end{tabular}
\end{table*}

\begin{table*}[]
\centering
\caption{Same as in Table \ref{table:ICME}, but only for shocks where both SIR and ICME were involved.}
\label{table:SIR_ICME}
\small
\begin{tabular}{@{}lllllllllll@{}}
\textbf{SIR/ICME shocks}  \\
\toprule
      & \textbf{Ions keV}      &            &            & \textbf{Ions MeV}   &            &            & \multicolumn{2}{l}{\textbf{Electrons keV} } &              &            \\
       \cmidrule(l){2-4}  \cmidrule(l){5-7}   \cmidrule(l){8-11}  
           & Back-   & No shock-  & Shock-     & Back-    & No shock-  & Shock-     & Back-    & No shock-  & Contami- & Shock-     \\
                &  ground & associated & associated &  ground  & associated & associated &  ground  & associated & nation   & associated \\
                &               & increase   & increase   &            & increase   & increase   &               & increase   &          & increase   \\
  \midrule
\textbf{STA}    & & & & & & & & & & \\
2007     & 0    & 1    & 0    & 1    & 0    & 0    & 1    & 0    & 0    & 0    \\
2008     & 0    & 0    & 0    & 0    & 0    & 0    & 0    & 0    & 0    & 0    \\
2009     & 0    & 0    & 0    & 0    & 0    & 0    & 0    & 0    & 0    & 0    \\
2010     & 0    & 0    & 1    & 1    & 0    & 0    & 0    & 0    & 1    & 0    \\
2011     & 0    & 0    & 0    & 0    & 0    & 0    & 0    & 0    & 0    & 0    \\
2012     & 0    & 1    & 1    & 1    & 0    & 1    & 0    & 2    & 0    & 0    \\
2013     & 0    & 3    & 0    & 3    & 0    & 0    & 0    & 2    & 0    & 1    \\
2014     & 0    & 4    & 1    & 2    & 3    & 0    & 0    & 5    & 0    & 0    \\
sum      & 0    & 9    & 3    & 8    & 3    & 1    & 1    & 9    & 1    & 1    \\
         &      &      &      &      &      &      &      &      &      &      \\
\textbf{STB}    &      &      &      &      &      &      &      &      &      &      \\
2007     & 0    & 2    & 0    & 2    & 0    & 0    & 2    & 0    & 0    & 0    \\
2008     & 0    & 0    & 0    & 0    & 0    & 0    & 0    & 0    & 0    & 0    \\
2009     & 1    & 0    & 0    & 1    & 0    & 0    & 1    & 0    & 0    & 0    \\
2010     & 0    & 0    & 1    & 1    & 0    & 0    & 0    & 0    & 1    & 0    \\
2011     & 0    & 3    & 0    & 3    & 0    & 0    & 0    & 3    & 0    & 0    \\
2012     & 0    & 1    & 3    & 4    & 0    & 0    & 0    & 3    & 0    & 1    \\
2013     & 0    & 0    & 0    & 0    & 0    & 0    & 0    & 0    & 0    & 0    \\
sum      & 1    & 6    & 4    & 11   & 0    & 0    & 3    & 6    & 1    & 1    \\
         &      &      &      &      &      &      &      &      &      &      \\
sum A\&B & 1    & 15   & 7    & 19   & 3    & 1    & 4    & 15   & 2    & 2    \\
av. M    & 2.00 & 1.39 & 1.88 & 1.55 & 1.71 & 1.42 & 1.80 & 1.44 & 2.21 & 1.41 \\
av. $\theta_{Bn}$ & 29.0 & 71.9 & 60.1 & 69.9 & 44.9 & 66.5 & 58.1 & 82.4 & 61.6 & 84.4 \\
\bottomrule
\normalsize
\end{tabular}
\end{table*}

\begin{table*}[]
\centering
\caption{Same as in Table \ref{table:ICME}, but only for shocks where no association to SIR or ICME was listed in the shock catalogs.}
\label{table:other}
\small
\begin{tabular}{@{}lllllllllll@{}}
\textbf{Other shocks}  \\
\toprule
      & \textbf{Ions keV}      &            &            & \textbf{Ions MeV}   &            &            & \multicolumn{2}{l}{\textbf{Electrons keV} } &              &            \\
       \cmidrule(l){2-4}  \cmidrule(l){5-7}   \cmidrule(l){8-11}  
           & Back-   & No shock-  & Shock-     & Back-    & No shock-  & Shock-     & Back-    & No shock-  & Contami- & Shock-     \\
                &  ground & associated & associated &  ground  & associated & associated &  ground  & associated & nation   & associated \\
                &               & increase   & increase   &            & increase   & increase   &               & increase   &          & increase   \\
  \midrule
\textbf{STA}    & & & & & & & & & & \\
2007     & 0    & 0    & 0    & 0    & 0    & 0    & 0    & 0    & 0    & 0 \\
2008     & 2    & 0    & 0    & 2    & 0    & 0    & 1    & 1    & 0    & 0 \\
2009     & 1    & 2    & 0    & 3    & 0    & 0    & 2    & 0    & 1    & 0 \\
2010     & 0    & 1    & 0    & 1    & 0    & 0    & 0    & 1    & 0    & 0 \\
2011     & 0    & 3    & 1    & 2    & 1    & 1    & 0    & 4    & 0    & 0 \\
2012     & 2    & 11   & 0    & 10   & 3    & 0    & 1    & 12   & 0    & 0 \\
2013     & 0    & 7    & 0    & 5    & 2    & 0    & 0    & 7    & 0    & 0 \\
2014     & 0    & 3    & 0    & 2    & 1    & 0    & 0    & 3    & 0    & 0 \\
sum      & 5    & 27   & 1    & 25   & 7    & 1    & 4    & 28   & 1    & 0 \\
         &      &      &      &      &      &      &      &      &      &   \\
\textbf{STB}    &      &      &      &      &      &      &      &      &      &   \\
2007     & 0    & 0    & 0    & 0    & 0    & 0    & 0    & 0    & 0    & 0 \\
2008     & 0    & 1    & 0    & 1    & 0    & 0    & 0    & 1    & 0    & 0 \\
2009     & 1    & 1    & 0    & 2    & 0    & 0    & 2    & 0    & 0    & 0 \\
2010     & 2    & 1    & 0    & 3    & 0    & 0    & 1    & 2    & 0    & 0 \\
2011     & 1    & 2    & 1    & 2    & 2    & 0    & 1    & 3    & 0    & 0 \\
2012     & 0    & 2    & 0    & 2    & 0    & 0    & 0    & 2    & 0    & 0 \\
2013     & 0    & 5    & 0    & 4    & 1    & 0    & 1    & 4    & 0    & 0 \\
sum      & 4    & 12   & 1    & 14   & 3    & 0    & 5    & 12   & 0    & 0 \\
         &      &      &      &      &      &      &      &      &      &   \\
sum A\&B & 9    & 39   & 2    & 39   & 10   & 1    & 9    & 40   & 1    & 0 \\
av. M    & 1.37 & 1.33 & 2.01 & 1.36 & 1.39 & 1.22 & 1.35 & 1.36 & 1.35 & 0 \\
av. $\theta_{Bn}$ & 54.8 & 53.2 & 76.5 & 56.6 & 43.7 & 76.8 & 47.6 & 55.8 & 61.2 & 0 \\
\bottomrule
\normalsize
\end{tabular}
\end{table*}

\end{appendix}
\end{document}